\documentclass{ws-p8-50x6-00}

\begin{document}

\title{Two-Loop Jet Physics: Status and Prospects
\footnote{{\rm Talk presented at 
9th International Workshop on ``Deep Inelastic Scattering and QCD'' 
(DIS 2001), April 2001, Bologna, Italy}}
}

\author{T.\ Gehrmann}

\address{Theory Division, CERN, CH-1211 Geneva 23, Switzerland}

\maketitle

\abstracts{I review the recent theoretical progress towards the 
computation of jet observables at two loops in QCD.}

\vspace{-5cm} \begin{flushright}CERN-TH/2001-179 
\end{flushright}\vspace{4cm}

\section{Introduction}
Jet production observables are among the most sensitive probes of QCD at
high energy colliders, where they are used for example to determine the
strong coupling constant. At present, the interpretation of jet production
data within perturbative QCD is restricted to next-to-leading 
order (NLO)
calculations, with theoretical uncertainties considerably larger
than current experimental errors. The extension of jet calculations to
NNLO requires various ingredients, such as two-loop corrections to
multi-leg amplitudes, multiple collinear limits of tree amplitudes, as
well as a consistent method for the numerical computation of jet
observables from NNLO parton level cross sections. 
In this talk, I review recent
progress made on these subjects. 

Jet observables of particular phenomenological relevance correspond to 
$2\to 2$ scattering or $1\to 3$ decay kinematics. They are:
\begin{itemize}
\item Three-jet observables in $e^+e^-$
\item DIS (2+1)-jet production 
\item Hadron-hadron  2-jet and $(V+1)$-jet production
\end{itemize}
The experimental  accuracy on these processes at LEP, HERA and the 
Tevatron has already reached a level that makes theoretical predictions 
beyond NLO desirable.

\section{Structure of NNLO jet physics}
At leading order in perturbation theory, a jet is approximated by a
single parton. Each parton in an event is required to be well separated 
in phase space from all other partons. Extending jet calculations to higher
orders, this simple picture is no longer true.
\begin{figure}[t]
\begin{center}
\begin{tabular}{|c|c|c|}\hline
{ Subprocess}&\parbox{2.8cm}{\vspace{0.1cm}
{ Partonic\\ final state}\vspace{0.1cm}} 
&\hspace{0.3cm}\parbox{2.2cm}{\vspace{0.1cm}
{ Partons\\ in jets}}\\
\hline
\parbox{4cm}{\vspace{0.1cm}
{  $\gamma^*\to 3$~partons, 2 loop}\\
{ e.g.\ }\vspace{-0.2cm} \\
\phantom{x}\hspace{0.4cm}~\epsfig{file=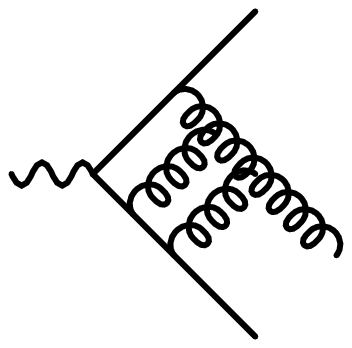,width=1.6cm}\vspace{0.1cm}
}
&\parbox{2.8cm}{  3 partons} &\parbox{1.6cm}{  (1) (1) (1)}
\\ \hline
\parbox{4cm}{\vspace{0.1cm}
{  $\gamma^*\to 4$~partons, 1 loop}\\
{ e.g.\ }\vspace{-0.2cm} \\
\phantom{x}\hspace{0.4cm}~\epsfig{file=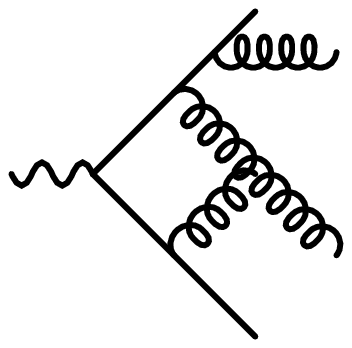,width=1.6cm}\vspace{0.1cm}
}
&\parbox{2.8cm}{
  {  4 partons\\ 
(3}{  +1}{  ) partons}}
&\parbox{1.6cm}{
{ (2) (1) (1)\\
 (1) (1) (1)}}
\\ \hline
\parbox{4cm}{\vspace{0.1cm}
{  $\gamma^*\to 5$~partons, tree}\\
{ e.g.\ }\vspace{-0.2cm} \\
\phantom{x}\hspace{0.4cm}~\epsfig{file=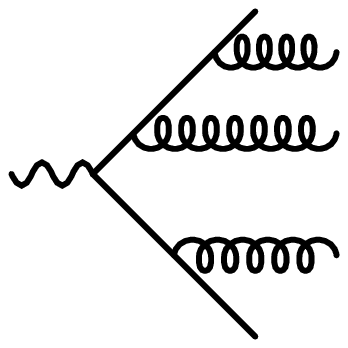,width=1.6cm}\vspace{0.1cm}
}
&\parbox{2.8cm}{
{  5 partons\\
\phantom{partons}\\
(4}{  +1}{  ) partons\\
(3}{  +2}{  ) partons
}}&\parbox{1.6cm}{{
  (3) (1) (1)\\
(2) (2) (1)\\
(2) (1) (1)\\
(1) (1) (1)}}
\\ \hline
\end{tabular}
\end{center}
\caption{Contributions to $\gamma^* \to 3$ jets at NNLO; $(m+n)$ partons
indicate $m$ theoretically resolved and $n$ theoretically unresolved 
(soft or collinear) partons.}
\label{fig:struc}
\vspace{-6mm}
\end{figure}

At NLO, one has to include one-loop virtual 
corrections to the leading order subprocess as well as corrections from 
one-particle real emission. Both these contributions contain divergences, 
which have to be extracted, usually using dimensional 
regularization~\cite{dreg,hv},
before a jet algorithm is applied to the different
final states in order to compute a particular jet observable. Since the 
jet algorithm acts differently on each partonic final state, it is not 
possible to combine contributions from real and virtual emission, as is 
done in multi-loop calculations of fully inclusive observables. 
It is for this reason that  calculations of jet observables 
were limited to NLO accuracy up to now. 

At the next-to-next-to-leading order (NNLO), one has to consider three 
different contributions: two-loop corrections to the leading order 
subprocess, one-loop corrections to the one-particle real emission, as 
well as two-particle real emission. 
Figure~\ref{fig:struc} shows the different contributions to three-jet 
production in $e^+e^-$ annihilation. All three contributions individually 
contain 
divergences  of infrared origin, which cancel only in the sum of the 
contributions. Again, each contribution has to be evaluated separately, 
since the (experimental) jet algorithm acts differently on final states with 
different numbers of partons.

\section{Real corrections}
Real corrections at NNLO consist of one-loop corrections to  one-particle 
emission processes and of two-particle emission processes. Since the 
infrared divergences of these real corrections have to be extracted before 
application of  the jet algorithm, it is necessary to introduce an 
infrared separation procedure, such as phase space slicing, subtraction, or 
a hybrid of these two methods. 

At present, general algorithms exist for the computation of 
the one-loop corrections to  one-particle 
emission processes~\cite{onel} both in the framework of 
slicing and of subtraction methods. The computation of 
all two-particle emission 
contributions relevant to jet physics still is, however, an 
open issue. Some of these contributions have been computed within the 
hybrid subtraction method in the context of the calculation of the 
photon+1~jet rate in $e^+e^-$ annihilation~\cite{gg}. This calculation 
is moreover the only example so far of a numerical 
implementation of NNLO corrections to a jet observable.

\section{Virtual corrections}
The jet production observables listed in the introduction correspond to 
$2\to 2$ scattering or $1\to 3$ decay kinematics, i.e.\ the 
relevant scattering amplitudes are four-point functions
with massless internal propagators and up to one external leg off shell. 
The large number of different integrals appearing in the 
corresponding two-loop Feynman amplitudes 
can be reduced to a small number of master integrals. 
The techniques used in these reductions are 
integration-by-parts identities~\cite{hv,chet} 
and Lorentz invariance~\cite{gr}. A computer algorithm for the 
automatic reduction of all two-loop four-point integrals 
has been derived~\cite{gr,laporta}. 

For two-loop four-point functions with massless internal propagators
and  all legs on shell, which are 
relevant for example in the NNLO
calculation of two-jet production at hadron colliders, all master
integrals have been calculated over the past two
years. The calculations~\cite{onshell1,onshell2} 
were performed using the Mellin--Barnes 
method and the differential equation 
technique~\cite{remiddi}. 
The resulting master integrals can be 
expressed in terms of Nielsen's generalized polylogarithms~\cite{nielsen}. 
Very recently, these master integrals were already applied in the 
calculation of 
two-loop virtual corrections to Bhabha scattering~\cite{m1},
 in the limit of vanishing 
electron mass, and to parton--parton scattering~\cite{m2}.
To obtain the full virtual corrections, one also needs to know the 
corresponding squared one-loop amplitudes~\cite{m3}.

We have used~\cite{gr1} the differential equation approach 
to compute all master integrals for two-loop four-point functions 
with one off-shell leg. Earlier partial results on these functions were 
available in the literature~\cite{glover,smirnew}, 
as well as  a purely numerical approach to 
their computation~\cite{num}.
We find full agreement with these earlier results. 
Our results~\cite{gr1} for these master
integrals are in terms of two-dimensional 
harmonic polylogarithms (2dHPL),
a generalization of the harmonic 
polylogarithms~\cite{hpl}. 
All 2dHPL appearing in the divergent parts of 
the master integrals can be expressed in terms of 
Nielsen's generalized polylogarithms of suitable non-simple arguments, while 
the 2dHPL appearing in the finite parts are one-dimensional integrals over 
generalized polylogarithms. An efficient numerical implementation of these 
functions is currently being worked out. Our results
correspond to the kinematical situation of a $1\to 3$ decay, their analytic 
continuation into the region of $2\to 2$ scattering processes requires 
the analytic continuation of the 2dHPL, which is outlined in~\cite{gr1}.  

These four-point two-loop master integrals 
with one leg off shell are a crucial ingredient to 
 the virtual NNLO  corrections to
 three-jet production in $e^+e^-$ annihilation, 
two-plus-one-jet production in $ep$ scattering and 
vector-boson-plus-jet production at hadron colliders. 

\section{Summary and Outlook}

Considerable progress towards the computation of jet observables has 
been made recently, concerning in particular the analytic calculation of
the virtual two-loop corrections to four-point amplitudes: 
all  two-loop amplitudes for parton--parton scattering are now known, 
and  all master integrals relevant to the decay of a vector boson 
into three partons have been calculated.

However, these 
corrections form  only part of a full NNLO 
calculation, which also has to include the one-loop corrections 
to processes with one soft or collinear real parton
 as well as 
tree-level processes with two soft or collinear partons.
While a general procedure exists for the calculation of the former, 
considerable effort is still needed to derive the latter. 

Only after summing all these contributions (and including 
terms from the renormalization of parton distributions 
for processes with partons in the initial state) do
the divergent terms cancel 
 one another. The remaining finite terms have to be combined 
into a numerical programme implementing the 
experimental definition of jet observables and event-shape variables. 
\section*{Acknowledgement}
I wish to thank 
E.\ Remiddi for a pleasant and fruitful collaboration on the topics 
discussed in this talk.

\end{document}